\documentclass[preprint,12pt]{elsarticle}

\usepackage{amssymb}

\journal{Journal of Magnetism and Magnetic Materials}

\begin{document}

\begin{frontmatter}



\title{Electric bias-controlled switching of magnetization of ferrimagnetically coupled Mn delta-layers in a GaAs-AlGaAs quantum well}

\author{N.\,V.\,Agrinskaya, A.\,M.\,Kalashnikova, V.\,I.\,Kozub}

\address{Ioffe Institute, 194021 St. Petersburg, Russia}

\begin{abstract}
We suggest a model of synthetic ferrimagnetic semiconductor structure based
on GaAs-AlGaAs quantum well doped by two Mn delta-layers. The coupling between the delta-layers is mediated by extra holes, and can be
switched between ferro- and antiferromagnetic one by gating the structure.
A proper choice of Mn concentrations in the delta-layers and of local degree of
disorder enables fabrication of a ferrimagnetic structure supporting ultrafast
switching of magnetization by short pulses of electric bias without an external magnetic field.
The switching mechanism in the structure relies on kinetic spin exchange between
the two delta-layers which is mediated by exchange scattering of electric-pulse heated holes by
magnetic ions within the layers. Owing to specific interplay between characteristics of the exchange scattering, spin decay times, and the heat withdraw in the suggested synthetic ferrimagnetic semiconductor, the necessary parameters of electric-bias pulse are within the technologically accessible range, and do not contradict typical thermal kinetics of semiconductor structures.
\end{abstract}

\begin{highlights}
\item We suggest a design of an artificial ferrimagnet consisting of two Mn-doped delta-layers with different concentrations placed in a GaAs-AlGaAs quantum well and coupled via extra holes in the well.
\item We analyze a scenario of the magnetization switching in the artificial ferrimagnet realized via heating the holes by a picosecond electric bias pulse which facilitates exchange scattering of the holes on Mn ions.
\end{highlights}


\end{frontmatter}

\section{Introduction}

Recently, a great attention \cite{Kirilyuk-RPP2013} was attracted by experiments demonstrating extremely fast ($\sim 10^{-12}$\,s)
magnetization reversal triggered by a single femtosecond laser pulse in a ferrimagnetic metallic rare-earth (RE) - transition metal
(TM) alloy GdFeCo without external magnetic field \cite{Stanciu-PRL2007,Vahaplar-PRL2009,Ostler-NComm2012}. Follow-up studies have suggested that laser-induced magnetization switching can be also realized in other RE-TM alloys, as well as in a variety of the engineered
ferrimagnetic structures, including exchange coupled RE-TM multilayers and heterostructures
comprised by two transition metal layers antiferromagnetically coupled through a nonmagnetic metallic
interlayer \cite{Mangin-NatureMat2014,Zhang-MPL2018,Lalieu-NComm2019}. Most importantly, experimental studies have demonstrated that the
all-optical reversal of magnetization is not precessional and
relies on subpicosecond quenching of the magnetizations \cite{Beaurepaire-PRL1996} of
RE and TM sublattices \cite{Vahaplar-PRL2009}. The time-resolved X-ray \cite{Radu-Nature2011,Radu-SPIN2015} and optical \cite{Khorsand-PRL2013} experiments have unveiled unconventional distinct dynamics of these sublattices leading to emergence of a transient ferromagnetic-like state. Such nonequilibrium
dynamics is believed to enable deterministic magnetization reversal, without any need for any other stimulus defining the magnetization
direction \cite{Ostler-NComm2012,Khorsand-PRL2012}.

Naturally, microscopical mechanism underlying unconventional
response of magnetization of a ferrimagnetic metallic system to a
femtosecond laser pulse is the subject of intense discussions
nowadays. Several theoretical and modelling approaches were employed to account for main features of the all-optical switching. Atomistic and multiscale calculations based on a Landau-Lifshitz-Bloch equation \cite{Garanin-PRB1997} for an
ensemble of the exchange-coupled spins were developed in Refs.\,\cite{Ostler-NComm2012,Vahaplar-PRB2012,Atxitia-PRB2013,Atxitia-PRB2014,Moreno-PRB2019} for describing the all-optical reversal in single phase alloys, as well as in exchange-coupled multilayers \cite{Evans-APL2014}, and to examine a feasibility of switching in ordered RE-TM alloys \cite{Moreno-PRB2019}. In \cite{Kimel} a comprehensive phenomenological model
based on the Onsager's relations \cite{Bar} was developed introducing an exchange-dominated regime of laser-induced dynamics in a ferrimagnet, which allows the reversal of magnetization solely due to the ultrafast heating. This work highlighted importance of the angular momentum exchange between the
sublattices of the ferrimagnet. In order to gain insight into microscopical nature of the all-optical switching, dissipationless energy and angular momentum exchange between TM and RE sublattices mediated by 5$d$-4$f$ coupling in RE ions has been explored in \cite{Wienholdt-PRB2013}, and the exchange electron-electron scattering as the driving mechanism of the magnetization reversal was discussed in \cite{Baral}. In \cite{Kalash} a general microscopic approach based on the rate equations was suggested for addressing the problem of the angular momentum exchange between two nonequivalent magnetic sublattices in a metal. In the latter work, the exchange scattering was found to be the driving mechanism of the switching. Similarly, switching enabled by the exchange scattering was also considered in \cite{Gridnev-PRB2018}, with however, principally different model of a RE-TM system. Importantly, since a laser pulse serves as an ultrafast heating pulse only \cite{Ostler-NComm2012,Khorsand-PRL2012}, in \cite{Kalash} it was suggested to realize electric-bias induced switching in a metallic ferrimagnetic structure. Recently, the first report appeared on the experimental observation of the switching of the magnetization of the GdFeCo wire by a picosecond current pulse \cite{Bokor}.

Along with unveiling a nature of unconventional dynamics of the spin system, the goal of these theoretical studies is to provide recipes for novel systems and alternative stimuli which enable ultrafast switching of magnetization under technologically accessible conditions \cite{Evans-APL2014,Moreno-PRB2019,Atxitia-APL2018}. Here we propose a synthetic semiconducting ferrimagnet where the exchange scattering-based magnetization switching can be realized upon application of an electric-bias pulse of a picosecond duration. The ferrimagnet is comprised by two ferromagnetic Mn-doped delta-layers in a GaAs-AlGaAs quantum well (QW). Antiferromagnetic coupling between the delta-layers is supported by the extra holes in the QW, and can be controlled by proper gating the structure. Ferrimagnetism of the whole structure is realized due to different Mn concentration in the ferromagnetic delta-layers. Using the general microscopic approach developed earlier in \cite{Kalash}, we show that ferrimagnetic properties of this structure support switching of the net magnetization via transient ferromagnetic-like state in zero applied magnetic field. Occurrence of the transient ferromagnetic-like state and consequent switching of the net magnetization of the structure is enabled by different magnetizations and Curie temperatures of the delta-layers, while the rates of exchange scattering of free carriers (holes) at the magnetic ions (Mn) are similar in both layers. The latter is in contrast to the all-optical or electric-bias switching in RE-TM alloys, where different exchange scattering rates for the TM and RE metals were playing a decisive role. Further, we argue that semiconducting properties make the switching more energetically profitable than in metals, and also ease constrains regarding short duration of the electric bias pulse required for switching.

\section{A synthetic ferrimagnet based on two Mn delta-layers in a GaAs-AlGaAs QW}

\begin{figure}[h]
    \centering
    \includegraphics[width=8.6cm]{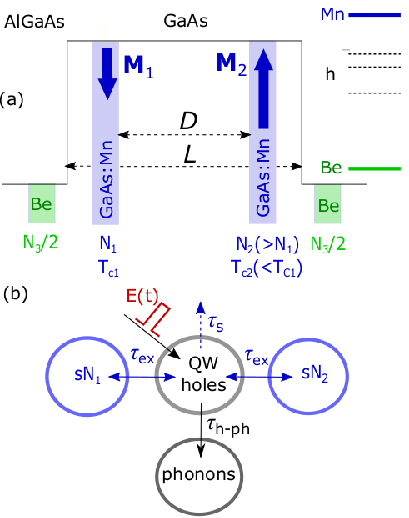}
    \caption{(a) Schematic picture of the ferrimagnetic structure comprised by two Mn-doped delta-layers placed within the GaAs-AlGaAs QW. Barrier is doped with shallow Be acceptor. Also shown are the Mn (blue line) and Be levels (green line), and laterally quantized hole levels (black lines). (b) Schematic diagram showing the spin and heat withdraw exchange channels considered in the model: blue lines show the exchange scattering and the spin relaxation channels, the black lines show the heating of the QW holes by the electric field $E(t)$ pulse and its cooling down via the phonon system.}
    \label{fig:1}
\end{figure}

As a candidate structure for a synthetic ferrimagnet we consider a QW GaAs-GaAlAs containing two delta-layers of Mn with concentrations $N_1$ and $N_2$. The delta-layers are separated by a distance $D$, which is comparable to the
well width $L\sim$15\,nm (Fig.\ref{fig:1}). Recently a system comprised by a Mn delta-layer in GaAs-based QW attracted attention \cite{Nazmul-PRL2005,Aronzon} as an alternative to a bulk diluted ferromagnetic semiconductor GaMnAs \cite{Dietl-RMP2014}.
These studies imply that the most promising realization of such a ferromagnet is the one with a delta-layer situated in the barrier in the vicinity of the QW. The ferromagnetic ordering is then realized due to indirect exchange supported by holes within
the well. It was suggested that in such a way the holes within the
well do not experience a disorder imposed by Mn delta-layer \cite{Nazmul-PRL2005,Aronzon}. This supposedly allows obtaining higher Curie temperatures $T_C$ despite of the tunneling exponent which is necessary to pay for the holes to contact Mn ions.

Recently an observation of ferromagnetism in delta-doped GaAs-AlGaAs QWs with Curie temperatures around 200\,K at unusually small
Mn doping levels of $\sim 5 \cdot 10^{12}$\,cm$^{-2}$ was reported in \cite{Agrin1,Agrin2} The indirect
exchange in the delta-layer in this case was supported by the holes supplied by Mn
atoms themselves. The lower concentration of Mn dopants lead to a
decrease of disorder potential which favors higher $T_C$ due to, in particular, a suppression of concentration
of Mn interstitials which are known to be compensating defects.
This fact was proved experimentally by demonstrating that an increase of Mn
concentration leads to a suppression of ferromagnetism in the in delta-doped GaAs-AlGaAs QWs \cite{Agrin1}.
This result advocates attempts to fabricate ferromagnetic structures by doping a region within a QW, which magnetic properties are controlled by holes within the well. Then, one can also fabricate {\it two} delta-layers of Mn in the same QW. Importantly, in this case the layers will be coupled to the same holes subsystem localized within the well.

Here we consider a particular case when $N_2 > N_1$, with $N_1$ being large enough to support ferromagnetism in the corresponding delta-layer. The separation distance $D$ between the delta-layers is much larger than the localization length for a Mn hole. Then the ferromagnetism of each delta-layer is supported by own holes of Mn acceptors \cite{Agrin1}. In this case the layer with Mn concentration $N_2$ is characterized by a lower value of
coupling between the ferromagnetic ions due to an increase of degree of disorder. The rough estimate of the disorder effect on $T_{C2}$ can be given by $T_{C2} \propto \exp (- N_2^{-1/2}/l)$. Here $l$ is the mean free
path for the holes in the vicinity of the second layer, and it is somewhat less than the distance between the Mn atoms $\sim
N_2^{-1/2}$. Note that the condition $l<N_2^{-1/2}$ can be also reached
by any additional intended contamination of the corresponding
interface region. In any case one can control the
value of $T_{c2}$ for a given value of $N_2 > N_1$. Thus we consider the system of two ferromagnetic layers 1 and 2 with $T_{c1}>T_{c2}$ and the saturation magnetizations $M_{s1}<M_{s2}$ (Fig.\,\ref{fig:1}).

Under the assumptions considered above, the delta-layers are not exchange coupled. An exchange coupling of a Ruderman–Kittel–Kasuya–Yosida (RKKY) type between the layers can be realized by additional carriers supplied by doping of the barriers with shallow Be acceptors. Doping both barriers with nearly the same surface concentrations $N_3/2$ allows cancelling the electric fields produced by the barrier acceptor
layers within the well. Thus, these fields do not affect the bare well potential.
The RKKY exchange coupling between the delta-layers could be either ferromagnetic or antiferromagnetic, depending on the band holes energies.
We note that additional holes within the well would improve ferromagnetic ordering in the
delta-layers via RKKY exchange coupling. We further note that a slightly higher doping level in the barrier closest
to the second layer can impose additional disorder potential on this
layer resulting from the screening the additional barrier charges by the holes within the second
layer.

Here we have to address the role of the second Hubbard band for the Mn acceptors, since
the Hubbard energy for them is at least twice lower than
ionization energy \cite{Aver}. Thus one expects that the
barrier dopants will initially lead to partial filling of the upper Hubbard
bands of the Mn acceptors. However, the complete filling of the upper
Hubbard band corresponds to additional hole and thus additional
positive charge per any site within the delta-layer. The concentration of Mn-sites allowing ferromagnetic ordering starts
around $10^{13}$ cm$^{-2}$ giving the intersite distances of order of 3\,nm \cite{Agrin2}, and the complete filling of the upper Hubbard
bands leads to large increase of the on-site hole energies resulting from coupling to the holes on the
neighboring sites (at distances at least of the order of distance
to the negatively charged layers within the barriers). Such holes will be inevitably pushed to the band states which are
delocalized within the well. Indeed, any charged neighbor at the
distance 3\,nm gives additional energy of the order of 30\,meV $\sim
300$\,K while the ionization energy for Mn is 100\,meV.
Correspondingly, only relatively small part of the upper Hubbard band can be situated lower than the bottom of the band resulting from lateral quantization. At partial occupations of the upper Hubbard bands the states of these bands are expected to be delocalized along the corresponding layer giving additional support to the ferromagnetic ordering within the layer. For larger concentration of the additional holes within the well, $N_3 - (N_1 + N_2) > \beta N_1$ ($\beta << 1$), the presence of the upper Hubbard band can be neglected due to a presence of the band holes. As a result, the whole system can be considered as the ferromagnetically ordered delta-layers with the delocalized band holes.

Concentration of these holes of the order of $10^{19}$ cm$^{-3}$ gives the total surface concentration of the
holes of the order of $10^{13}$\,cm$^{-2}$ \cite{Agrin2}, that is of the order of the surface concentration within the ferromagnetic layers. In this case the screening length is of the order of $\sim 2$\,nm for the holes with energy $\sim30$\,meV ($\sim 300 $K). Such distance is smaller or, at least, comparable to the intersite distances within the layers. Thus, at corresponding concentrations and the free holes energies we can consider the potentials of the doubly occupied sites to be screened. This gives and additional argument to
neglect a filling of the upper Hubbard band in our model.

We assume $N_1 \sim 5 \cdot 10^{12}$ cm$^-2$, $N_2 \sim  1.5 \cdot 10^{13}$ cm$^{-2}$, $N_3 \sim  10^{13}$ cm$^{-2}$. For
the width of the well $L \sim 1.5\cdot 10^{-6}$ cm and the hole mass
$10^{-27}$\,g one estimates the energy of the first quantization level of
$\hbar^2 \pi^2/2mL^2 \sim 2 $\,meV, the Fermi
wavelength of the order of $10^{-6}$ cm and concentration of the
order of $10^{12}$ cm$^{-2}$. For concentration of the order of
$10^{13}$ cm$^{-2}$ distributed over the well one has the
interhole spacing $\sim 10^{1/2} \cdot 10^{-7}$ cm that is around
$\lambda_F \sim $ 3 nm. It corresponds to occupation of the 3$d$
level of lateral quantization which gives oscillating wave
function providing a possibility to have RKKY exchange coupling of any sign between the Mn delta-layers. Here we also note that the Fermi energy of
the band holes at the considered concentrations becomes to be
of the order or larger than the temperature of the system (200 - 300\,K). Thus, at the temperatures close to the
critical ones, which are also around 200 - 300 K, the situation is
close to degeneration which facilitates the RKKY interaction.

We note that several attempts to realize and study exchange
interactions between two Mn delta-layers imbedded into GaAs matrix were reported (see
e.g.  \cite{Hong-PRL2005,Menshov-PRB2008}). However the main attention in these studies has been paid to the role of
the carriers supplied by Mn itself, and the interlayer
distance was assumed to be small in comparison to localization
length of the Mn holes in the GaAs host. This is in strong contrast to the structure considered in the present work, where the additional holes are supplied by dopants within the AlGaAs barriers. As a result the distance between Mn layers can be at
least of the order of several localization lengths, and the wavelength of
the doped holes is comparable to the interlayer distance, and the
situation is clearly corresponds to the RKKY interactions. Importantly, this interlayer RKKY exchange coupling can be controlled by the
gate situated in any of the barriers. Indeed, the gate potential allows to control the concentration of the band holes within the
well, which, in turn, affects the character of interlayer coupling.

\section{Switching the magnetization of the delta-layers by electric-bias pulse}

Now we consider an evolution of the magnetizations of the two delta-layers with antiferromagnetic RKKY exchange between them in response to a short electric bias pulse of a duration of a few picoseconds. The application of the electric bias pulse rapidly increases the temperature of the holes in the QW. To treat the following evolution of the spin system we use the model developed in \cite{Kalash}. In brief, it describes evolution of the occupation numbers of two \textit{different} ferromagnetic sublattices coupled antiferromagnetically via delocalized carriers. The spin exchange between the localized ferromagnetic subsystems is mediated by delocalized carriers which temperature is rapidly increased above critical temperatures of both magnetic sublattices. Either femtosecond laser excitation or applying short electric bias pulse can serve for the carriers heating. Spin exchange results in the switching of the net magnetization of the system without any additional stimuli, such as external magnetic field, circular polarization of light, or spin polarization of a current. The switching of the net magnetization in this model relies on a delicate balance between the exchange scattering, spin relaxation, and cooling times. Importantly, this model is not restricted to the case of RE-TM alloys or heterostructures, and is also applicable for the case of the structures composed by two different transition metals.

The important difference between the system analyzed here and the RE-TM metallic systems considered in \cite{Kalash}, is that here two ferromagnetic subsystems, i.e. two delta-layers, are comprised by the same Mn ions. Thus, the exchange scattering of the holes on Mn ions is nearly equal for the both subsystems. We can estimate the characteristic times of this scattering in the following way. A 3D cross-section for Coulomb scattering of the mobile holes by an unscreened ion is of $(e^2/\kappa \varepsilon)^2 \sim 10^{-12}$ cm$^2$. 3D geometry is considered since the holes are delocalized in the direction normal to the well plane.  However, the effect of nonlinear screening by the holes, in particular by the ones supplied by Mn atoms itself, diminish this estimate down to $10^{-13}$ cm$^2$ which is controlled by the distance between Mn atoms. The cross-section for the exchange scattering is expected to be less by a factor $\gamma^2$, where $\gamma$ characterizes the relative strength of the exchange potential, and can be estimated to be of $\sim0.3$. Now we take into account that the inverse mean free path with respect to the exchange scattering is given by a product of this cross-section and 3D concentration of Mn ions in the well ($\sim 10 ^{19}$ cm$^{-3}$). This gives the value of $10^{-5}$\,cm for the mean free path. Correspondingly, for the holes velocity $\sim 10^7$ cm/s \cite{Dalal-JAP1971} the exchange scattering time can be estimated to be of $\tau_\mathrm{ex}\sim$1\,ps.

Another important difference between the spin dynamics in the considered artificial ferrimagnet and in the RE-TM metallic alloy is that the spin relaxation times in the semiconductor GaAs appear to be considerably longer than the exchange scattering times. In \cite{Koren} the spin relaxation time for the holes in GaAs structures was found to be of $\sim$20\,ps. This value is significantly larger than the spin relaxation time $\sim 0.1 - 1$\,ps in metals. Therefore, at the first stage of the electric-bias driven evolution of the spin balance in the system, the effect of spin relaxation can be neglected. We note that this assumptions corresponds to the model suggested in \cite{Gridnev-PRB2018}, where the two ferromagnetic subsystems with different magnetizations and similar exchange scattering times were considered. As we discussed above, such assumptions are justified in the case of Mn layers within the semiconductor QW, while their applicability to the metallic RE-TM alloy is arguable.

Right after the application of the electric bias pulse, the temperature of holes in the QW is increased to the value $T_h^*$, and the evolution of the spin system at this stage is reduced to redistribution of the total angular momentum $s(N_2 - N_1)$, where $s$ is the spin of a hole, between the two ferromagnetic subsystems. When the holes temperature reached $T_h^*$, the resulting momenta $sn_1$ and $sn_2$ of the layers are then given by
\begin{equation}
n_1 = \frac{(N_2 - N_1)}{N_2 + N_1} N_1,\hskip1cm n_2 = \frac{(N_2 - N_1)}{N_2 + N_1}N_2.\label{Eq:n1n2}
\end{equation}
As it is seen from Eqs.\,(\ref{Eq:n1n2}), the sign of the magnetization for the both subsystems now coincide since we consider the system with $N_2 > N_1$. That is at this stage the mutual spin orientations in the two delta-layers correspond to a ferromagnetic-like state.\cite{Radu-Nature2011}

Evolution of delta-layers magnetizations from their equilibrium values to the ones given by Eqs.\,(\ref{Eq:n1n2}) occurs on the time scale given by the estimated exchange scattering time $\tau_\mathrm{ex}\sim$1\,ps. We note that, in contrast to the evolution of the TM and RE magnetizations in the metallic RE-Tm alloys \cite{Kalash}, the particular time when the $M_{s1}$ vanishes appear not to play crucial role, since the spin relaxation times are much slower that the exchange scattering times. Only if the spin relaxation is fast or if the temperature of the system remains elevated (i.e. no cooling is present) then the total angular moment, and, correspondingly, the total magnetization would finally vanish.

To describe the evolution of the system from the ferromagnetic-like state we take into account the cooling down process. First, we consider a case when the dominant cooling mechanism is related to optical phonons with a frequency $\omega_0$. This is relevant when the temperature of the holes is high enough.
In this case the cooling time necessary to lower the temperature of the holes from $T_h^*$, down to some value $T_h$, can be roughly estimated as $t = \tau_\mathrm{h-ph}k_B(T_h^* - T_h)/(\hbar \omega_0)$, where the characteristic hole-phonon relaxation time is $\tau_\mathrm{h-ph} \sim 10^{-12}$\,s \cite{Cardona-book,Scholz-JAP1995}.

At temperatures lower that $\hbar \omega_0$ the further cooling is assisted by acoustical phonons. To estimate the corresponding relaxation time we take into account that the holes are coupled to phonons with momentum equal or less than the holes momentum due to a momentum conservation. The holes velocities are of $\sim10^7$ cm/s for temperature range in a vicinity of the Curie temperatures of the layers, i.e. in the range of 200 - 300\,K \cite{Dalal-JAP1971}. The wavevectors of corresponding phonons are less than $10^7$\,cm$^{-1}$. Thus, one concludes that the phonon energies $\hbar \omega$ are about 10 times lower than the holes energies $\varepsilon_h$. In this case the energy relaxation time is larger by a factor $\sim (\varepsilon_h/\hbar \omega)^2$ than the typical hole-phonon relaxation time $\tau_\mathbf{h-ph}\sim10^{-12}$\,s, provided that the temperature $T_h$ is not much below the room temperature. Thus, we obtain characteristic time $t = \tau_\mathbf{h-ph}(k_BT_h/\hbar \omega)^2$ for the further energy relaxation.

In our estimates we assume that the heat withdrawal from the quantum well is efficient enough, which can be ensured by the purity
of the AlGaAs barriers and small mismatch of the elastic constants between GaAs and AlGaAs. Thus, the phonon transport outside of the
well can be considered as a ballistic one, and the characteristic time of the phonon escape can be estimated as $D/w$, where $D$ is of the order of the thickness of the QW, and $w$ is the sound velocity. Than the phonon escape time is of the order of $10^{-12}$\,s which is comparable to the hole-phonon relaxation time $\tau_{h-ph}$. The heat capacity of phonon subsystem is much larger than that of the holes subsystem. For hole concentrations around $3 \cdot 10^{12}$ cm$^{-2}$ the corresponding factor is around $10^{4}$.
Thus at time scale less than $10^{-8}$\,s the problem of heat
withdrawal can be neglected.

Since the critical temperature for the first subsystem, $T_{c1}$ is larger, then that for the second subsystem
($T_{c2}$), the temperature of the holes reaches the value of $T_{c1}$ first upon cooling. Once the holes temperature is below $T_{c1}$ the magnetization of the first subsystem starts to restore according to standard thermodynamical law as (see e.g. \cite{Vons})
\begin{equation}
\Delta M_{s1} \sim  M_{s1} \left(\frac{T_{c1} - T_h}{T_{c1}}\right)^{1/2},
\end{equation}
where the $T_h$ is controlled by a process of cooling. Thus at a moment when
\begin{equation}
N_1\left(\frac{T_{c1} - T_h}{T_{c1}}\right)^{1/2} \sim \frac{(N_2 - N_1)}{N_2 + N_1}N_2
\end{equation}
the magnetization of the layer 1 starts to dominate over the magnetization of the layer 2 provided that it this moment the holes temperature is still larger than the critical temperature $T_{c2}$.

At the final stage of the process, i.e. when the holes temperature is lowered down to the critical temperature $T_{c2}$ of the
layer 2, the magnetization of the latter starts to restore as well. The preferential direction for the layer 2 magnetization is now set by the antiferromagnetic RKKY coupling to the magnetization of the layer 1. Correspondingly, the net magnetization of the system is switched at this stage.

\section{Conclusions}
We suggested a model of a synthetic ferrimagnetic structure based on the Mn-doped GaAs-AlGaAs quantum well. It contains two Mn
delta-layers within the well with different Mn concentrations and different degrees of disorder. Two symmetric layers of barrier acceptors are situated on different sides of the well in order to provide extra holes to the QW. These extra holes are essential since they participate in the RKKY coupling between the delta-layers. The sign of the RKKY coupling can be controlled by concentration of these holes, by distance between Mn layers, and by gating the structure. As a result, a structure of two antiferromagnetically coupled Mn delta-layers can be fabricated with one layer possessing larger saturation magnetization and lower critical temperature that the second layer. Using recently suggested theory for the switching of a ferrimagnet driven by the kinetic exchange scattering, we show that the magnetization of such semiconductor-based synthetic ferrimagnet can be switched by a short pulse of electric bias with no additional external stimuli, e.g. external magnetic field, required to set the magnetization direction.

We note that the suggested switching scenario of the synthetic semiconductor-based ferrimagne can be also realized by using short laser pulses instead of electric bias ones, since the former are known to trigger ultrafast demagnetization in GaMnAs \cite{Wang-PRB2008}.  It may be important from a point of view of integration of Mn-doped GaAs-AlGaAs QW allowing toggle switching of magnetization with the ultrashort semiconducor laser sources. GaAs and AlGaAs are among media for developing such lasers (for a review see, e.g., \cite{Tilma-Light2015} and more publications on this subject, e.g. \cite{Mayer-NComm2017}). Thus, one could envisage designing a device contained both a ultrafast laser source and the magnetic toggle switch "on a single chip". For the later to be realized, the Curie temperatures of the Mn-doped GaAs-AlGaAs QW should be above room temperature. Mn delta-doped GaAs possesses one of the highest temperatures among the III-V semiconductors reaching 250\,K \cite{Nazmul-PRL2005}. On the other hand, among the bulk ferromagnetic III-V semiconductors, those doped with Fe ions have the Curie temperature higher than (Ga,Mn)As \cite{Tu-PRB2015}. Therefore, it would be important to investigate if the III-V semiconductor-based structures with Fe delta-layers can be fabricated for the ultrafast switching. Presently, however, there are no reports on successful increase of the Curie temperature by Fe delta-doping of a III-V semiconductor \cite{Nishijima-JCrGr2019}.

\section{Acknowledgements}
N.V.A. and V.I.K. acknowledge financial support from the Russian Foundation for Basic Research, grant No.\,19-02-00184.


\begin{thebibliography}{00}

\bibitem{Kirilyuk-RPP2013} A. Kirilyuk, A. V. Kimel, Th. Rasing, Rep. Prog. Phys. \textbf{76}, 026501 (2013).
\bibitem{Stanciu-PRL2007} C. D. Stanciu, F. Hansteen, A. V. Kimel, A. Kirilyuk, A. Tsukamoto, A. Itoh, and Th. Rasing, Phys. Rev. Lett. \textbf{99}, 047601 (2007).
\bibitem{Vahaplar-PRL2009} K. Vahaplar, A. M. Kalashnikova, A. V. Kimel, D. Hinzke, U. Nowak, R. Chantrell, A. Tsukamoto, A. Itoh, A. Kirilyuk, and Th. Rasing, Phys. Rev. Lett. \textbf{103}, 117201 (2009).
\bibitem{Ostler-NComm2012} T. A. Ostler, J. Barker, R. F. L. Evans, R. W. Chantrell, U. Atxitia, O. Chubykalo-Fesenko, S. El Moussaoui, L. Le Guyader, E. Mengotti, L. J. Heyderman,    F. Nolting, A. Tsukamoto, A. Itoh, D. Afanasiev,  B. A. Ivanov, A. M. Kalashnikova, K. Vahaplar, J. Mentink, A. Kirilyuk, Th. Rasing, and A.V. Kimel, Nature Commun. \textbf{3}, 666 (2012).
\bibitem{Mangin-NatureMat2014} S. Mangin, M. Gottwald, C-H. Lambert, D. Steil, V. Uhlir, L. Pang, M. Hehn, S. Alebrand, M. Cinchetti, G. Malinowski, Y. Fainman, M. Aeschlimann and E. E. Fullerton, Nature Mater. {\bf 13} , 286 (2014).
\bibitem{Zhang-MPL2018} G. P. Zhang, M. Murakami, M. S. Si, Y. H. Bai, and T. F. George, Mod. Phys. Lett. B \textbf{32}, 1830003 (2018).
\bibitem{Beaurepaire-PRL1996} E. Beaurepaire, J.-C. Merle, A. Daunois, and J.-Y. Bigot, Phys. Rev. Lett. \textbf{76}, 4250 (1996).
\bibitem{Lalieu-NComm2019} M.L.M. Lalieu, R. Lavrijsen, B. Koopmans, Nature Commun. \textbf{10}, 110 (2019).
\bibitem{Radu-Nature2011} I. Radu,  K. Vahaplar,    C. Stamm,   T. Kachel,  N. Pontius, H. A. Dürr,    T. A. Ostler,   J. Barker,   R. F. L. Evans,    R. W. Chantrell, A. Tsukamoto, A. Itoh, A. Kirilyuk, Th. Rasing, A. V. Kimel, Nature \textbf{472}, 205 (2011).
\bibitem{Radu-SPIN2015} I. Radu, C. Stamm, A. Eschenlohr, F. Radu, R. Abrudan, K. Vahaplar, T. Kachel, N. Pontius, R. Mitzner, K. Holldack, A. Föhlisch, T. A. Ostler, J. H. Mentink, R. F. L. Evans, R. W. Chantrell, A. Tsukamoto, A. Itoh, A. Kirilyuk, A. V. Kimel and Th. Rasing, Ultrafast and Distinct Spin Dynamics in Magnetic Alloys, SPIN \textbf{5}, 1550004 (2015).
\bibitem{Khorsand-PRL2013} A. R. Khorsand, M. Savoini, A. Kirilyuk, A. V. Kimel, A. Tsukamoto, A. Itoh, and Th. Rasing, Element-Specific Probing of Ultrafast Spin Dynamics in Multisublattice Magnets with Visible Light, Phys. Rev. Lett. \textbf{110}, 107205 (2013).
\bibitem{Khorsand-PRL2012} A. R. Khorsand, M. Savoini, A. Kirilyuk, A. V. Kimel, A. Tsukamoto, A. Itoh, and Th. Rasing, Phys. Rev. Lett. \textbf{108}, 127205 (2012).
\bibitem{Garanin-PRB1997} D. A. Garanin, Phys. Rev. B \textbf{55}, 3050 (1997).
\bibitem{Vahaplar-PRB2012} K. Vahaplar, A. M. Kalashnikova, A. V. Kimel, S. Gerlach, D. Hinzke, U. Nowak, R. Chantrell, A. Tsukamoto, A. Itoh, A. Kirilyuk, and Th. Rasing, Phys. Rev. B \textbf{85}, 104402 (2012).
\bibitem{Atxitia-PRB2013} U. Atxitia, T. Ostler, J. Barker, R. F. L. Evans, R. W. Chantrell, and O. Chubykalo-Fesenko, Phys. Rev. B \textbf{87}, 224417 (2013).
\bibitem{Atxitia-PRB2014} U. Atxitia, J. Barker, R. W. Chantrell, and O. Chubykalo-Fesenko, Phys. Rev. B \textbf{89}, 224421 (2014).
\bibitem{Moreno-PRB2019} R. Moreno, S. Khmelevskyi, and O. Chubykalo-Fesenko, Phys. Rev. B \textbf{99}, 184401 (2019).
\bibitem{Evans-APL2014} R. F. L. Evans, Th. A. Ostler, R. W. Chantrell, I. Radu, and Th. Rasing, Appl. Phys. Lett. 104, 082410 (2014).
\bibitem{Kimel} J. H. Mentink, J. Hellsvik, D. V. Afanasiev, B. A. Ivanov, A. Kirilyuk, A. V. Kimel, O. Eriksson, M. I. Katsnelson, and Th. Rasing, Phys. Rev. Lett. {\bf 108}, 057202 (2012).
\bibitem{Bar} V. G. Baryakhtar, Zh. Eksp. Teor. Fiz. {\bf 87}, 1501 (1984); {\bf 94}, 196 (1988); Fiz. Nizk. Temp. {\bf 11}, 1198 (1985).[Sov. Phys. JETP {\bf 60}, 863 (1984); Sov. Phys. JETP {\bf 67}, 757 (1988); Sov. J. Low Temp. Phys. {\bf 11}, 662 (1985)].
\bibitem{Wienholdt-PRB2013} S. Wienholdt, D. Hinzke, K. Carva, P. M. Oppeneer, and U. Nowak, Phys. Rev. B \textbf{88}, 020406(R) (2013).
\bibitem{Baral} A. Baral, H. C. Schneider, Phys. Rev. B \textbf{91}, 100402 (2015).
\bibitem{Kalash} A. M. Kalashnikova and V. I. Kozub, Phys. Rev. B {\bf 93}, 054424 (2016).
\bibitem{Gridnev-PRB2018} V. N. Gridnev, Phys. Rev. B \textbf{98}, 014427 (2018).
\bibitem{Bokor} Y. Yang, R. B. Wilson, J. Gorchon, Ch.-H. Lambert, S. Salahuddin, and J. Bokor, Sci. Adv. \textbf{3}, e1603117 (2017).
\bibitem{Atxitia-APL2018} U. Atxitia and  T. A. Ostler, Appl. Phys. Lett. \textbf{113}, 062402 (2018).
\bibitem{Nazmul-PRL2005} A. M. Nazmul, T. Amemiya, Y. Shuto, S. Sugahara, and M. Tanaka, Phys. Rev. Lett. \textbf{95}, 017201 (2005).
\bibitem{Aronzon} B. A. Aronzon, M.A. Pankov, V.V. Rylkov, et al., J. Appl. Phys. \textbf{107}, 023905 (2010).
\bibitem{Dietl-RMP2014} T. Dietl and H. Ohno, Rev. Mod. Phys. \textbf{86}, 187 (2014).
\bibitem{Agrin1} N.V. Agrinskaya, V.A.Berezovets, A.Bouravlev , V.I.Kozub, Solid State Commun. {\bf 183}, 27 (2014).
\bibitem{Agrin2} N.V. Agrinskaya, V.A. Berezovets, V.I. Kozub, J. Magn. Magn. Mater. \textbf{466}, 180 (2018).
\bibitem{Aver} N.S.Averkiev, Yu.T.Rebane, I.N.Yassievich Sov. Phys. - Semiconductors, {\bf 19}, 98 (1988).
\bibitem{Hong-PRL2005} J. Hong, D.-Sh. Wang, and R. Q. Wu, Phys. Rev. Lett. \textbf{94}, 137206 (2005).
\bibitem{Menshov-PRB2008} V. N. Men’shov, V. V. Tugushev, P. M. Echenique, S. Caprara, and E. V. Chulkov, Phys. Rev. B \textbf{78}, 024438 (2008).
\bibitem{Koren} V.L. Korenev, I.A. Akimov, S.V. Zaitsev, V.F. Sapega, L. Langer, D.R. Yakovlev, Yu. A. Danilov, and M. Bayer, Nature Commun. \textbf{3}, 959 (2012).
\bibitem{Vons} S.V. Vonsovskii, Magnetism, J. Wiley (1974), ch. 18.
\bibitem{Wang-PRB2008} J. Wang, {\L}. Cywi\'{n}ski, C. Sun, J. Kono, H. Munekata, and L. J. Sham, Phys. Rev. B \textbf{77}, 235308 (2008).
\bibitem{Dalal-JAP1971} V. L. Dalal, A. B. Dreeben, and A. Triano, J. Appl. Phys. \textbf{42}, 2864 (1971).
\bibitem{Cardona-book} P. Yu, M. Cardona, \textit{Fundamentals of semiconductors} (4th ed., Springer, 2010), page 213.
\bibitem{Scholz-JAP1995} R. Scholz, J. Appl. Phys \textbf{77}, 3219 (1995).
\bibitem{Tilma-Light2015} B. W. Tilma, M. Mangold, Ch. A. Zaugg, S. M. Link, D. Waldburger, A. Klenner, A. S. Mayer, E. Gini, M. Golling, and U. Keller, Light: Sci. Appl. \textbf{4}, e310 (2015).
\bibitem{Mayer-NComm2017} B. Mayer, A. Regler, S. Sterzl, T. Stettner, G. Koblmuller, M. Kaniber, B. Lingnau, K. L\"{u}dge, and J. J. Finley, Nature Commun. \textbf{8}, 15521 (2017).
\bibitem{Tu-PRB2015} Nguyen Thanh Tu, Pham Nam Hai, Le Duc Anh, and M. Tanaka, Phys. Rev. B \textbf{92}, 144403 (2015).
\bibitem{Nishijima-JCrGr2019} Kento Nishijima, Nguyen Thanh Tu, Masaaki Tanaka, and Pham Nam Hai, J. Cryst. Growth 511, 127 (2019).

\end{thebibliography}
\end{document}